\def \mnras {MNRAS}

\documentclass{iau}
\usepackage{graphicx}

\title[Evolution of Massive Galaxy Structural Properties and Sizes via Star Formation]{Evolution of Massive Galaxy Structural Properties and Sizes via Star Formation}

\author[Jamie Ownsworth]{Jamie~R.~Ownsworth$^{1}$\thanks{E-mail: ppxjo1@nottingham.ac.uk}, Christopher~J.~Conselice$^1$, Alice~Mortlock$^{1}$,  William~G.~Hartley$^{1}$, Fernando~Buitrago$^{1,2}$ \\}

\affiliation{$^1$ University of Nottingham, School of Physics and Astronomy, Nottingham, NG7 2RD, U.K.  \\ 
$^2$ Institute for Astronomy, University of Edinburgh, Royal Observatory, Edinburgh, EH9 3HJ, U.K.} 

\pubyear{2012}
\volume{28}  
\pagerange{xx--xx}
\jname{Ultraviolet emission in early\--type galaxies}

\begin{document}

\maketitle

\begin{abstract}
We investigate the resolved star formation properties of a sample of 45 massive galaxies ($M_{*}>10^{11}M_{\odot}$) within a redshift range of $1.5 \le z \le 3$ detected in the GOODS NICMOS Survey (Conselice et al. 2011), a HST $H_{160}$\--band imaging program. We derive the star formation rate as a function of radius using rest frame UV data from deep $z_{850}$ ACS imaging. The star formation present at high redshift is then extrapolated to $z=0$, and we examine the stellar mass produced in individual regions within each galaxy. We also construct new stellar mass profiles of the in\--situ stellar mass at high redshift from S\'{e}rsic fits to rest-frame optical, $H_{160}$\--band, data. We combine the two stellar mass profiles to produce a modelled evolved stellar mass profile. We then fit a new S\'{e}rsic profile to the evolved profile, from which we examine what effect the resulting stellar mass distribution added via star formation has on the structure and size of each individual galaxy. We conclude that due to the lack of sufficient size growth and S\'{e}rsic evolution by star formation other mechanisms such as merging must contribute a large proportion to account for the observed structural evolution from $z>1$ to the present day.
\end{abstract}
In summary: We find three different profiles of star formation within the massive galaxies in this sample, Non\--significant Star Formation Growth (NG), Outer Star Formation Growth (OG) and Inner Star Formation Growth (IG) (see Ownsworth et al. 2012). With most of this sample of massive galaxies falling in to NG class using the derived tau model of evolution.We find that the star formation we observe at high redshift, and its effects on galaxy sizes, is not large enough to fully explain the observed size evolution of $\sim300\--500\%$. Star formation alone alone can only produce an increase in effective radius on the order of $\sim16\%$ over the whole sample. This value can vary as much a a factor of 4.5 by using different evolution mechanisms but is always insufficient to fully explain the observations. We find that over the whole sample of massive galaxies the stellar mass added via star formation has a slight effect on the surface brightness of the evolved galaxy profile such that their S\'{e}rsic indices decrease. This indicates that the star formation within these galaxies follows the same radial distribution as the original stellar mass profile. This also implies that star formation evolution has a minimal effect on structural evolution between $z\sim3$ and the present day. The increase in effective radius can be enhanced by adding in the effects of stellar migration to the stellar mass created via star formation. This increases the total effective radius growth to $\sim55\%$, which is still however much smaller than the total observed size increase.

\end{document}